\documentclass[12pt,a4paper]{article}

\usepackage{graphics}
\usepackage{latexsym}
\usepackage{amsmath}
\usepackage{amssymb}

\title{A simple model of
the hierarchical formation of galaxies}
\author{B. Hoeneisen}
\date{\small{Universidad San Francisco de Quito \\
	4 September 2000}}
\begin{document}
\maketitle

\begin{abstract}
\noindent
We develop a simple, fast and predictive model of the
hierarchical formation of galaxies which is
in quantitative agreement with observations. Comparing
simulations with observations we place
constraints on the density of the
universe and on the
power spectrum of density fluctuations.
\end{abstract}

\tableofcontents

\section{Introduction}
Simulations of galaxy formation usually start
with a set of gravitating point particles with given 
initial conditions which are then stepped forward
in time using huge computer resources. Here we
consider a complementary approach with a density 
distribution that is continuous instead of discrete.
The model is presented in Section 2 and
compared with observations in Section 3.
Conclusions are collected in Section 4.

\section{Model of galaxy formation}
\subsection{The hierarchical formation of galaxies}
Let us study linear departures from a critical
universe dominated by non-relativistic dark 
matter.\cite{Weinberg}
We consider a cube of side $L(t) = a_c(t)L_0$
and expand the density in this cube 
in a Fourier series. For growing
modes with $\delta_c \ll 1$ we define
the linear approximation of
the density of the universe as
follows:
\begin{eqnarray}
\rho_{lin}(\vec{x}, t, I) & \equiv &  
\frac{1}{6 \pi G t^2}\{1+\delta_c(\vec{x}, t)\} 
\equiv \frac{\rho'_{c0}}{a_c^3}
\{1+\delta_c(\vec{x}, t)\} \nonumber \\
& = & \frac{\rho'_{c0}}{a_c^3} 
\left\{ 1 + \Delta(\Omega_0)a_c + 
\sum_{\vec{k}, 0 < k \le k_I} \vert \delta_{\vec{k}} \vert a_c
\cdot \exp \left[ i \frac{\vec{k} \cdot \vec{x}}{a_c} + i
\varphi_{\vec{k}} \right] \right\}
\label{density}
\end{eqnarray}
where $t$ is the age of the universe, the subscript 
$0$ denotes present day values,
$\rho'_{c0} \equiv  \omega(\Omega_0) \rho_{c0}$, 
$\rho_{c0} \equiv 3 H_0^2 /(8 \pi G)$ is the critical density,
and $a_c(t) \equiv (t/t_0)^{2/3}$. 
$\varphi_{\vec{k}}$ are random phases.
Until Section 3.7 we will take the cosmological 
constant $\Lambda = 0$.
The functions $\Delta(\Omega_0)$ and
$\omega(\Omega_0) = [2/(3H_0 t_0)]^2$ are 
given in Tables \ref{Delta} and \ref{omega}.
The sum of the Fourier series is over comoving
wavevectors that satisfy periodic boundary conditions:
\begin{equation}
\vec{k} = \frac{2\pi}{L_0}(n\hat{\imath} + m\hat{\jmath} +
l\hat{k})
\label{k}
\end{equation}
where $n, m, l = 0, \pm 1, \pm 2, \pm 3 \cdots$ and
$1 \le [n^2 + m^2 + l^2]^{1/2} \le I$. The density in
Equation (\ref{density}) is due to Fourier components up
to wavevectors of magnitude $k_I \equiv 2 \pi I / L_0$.
Note that the critical density 
$\rho_c = \rho_{c0}/a_c^3$ is proportional to
$a_c^{-3}$, the density contrast $\delta_c$ 
grows in proportion
to $a_c$, and the \textquotedblleft{wavelengths}" 
$\lambda = 2\pi a_c/|\vec{k}|$ 
of the Fourier components stretch in proportion to $a_c$
(we use the word \textquotedblleft{wavelength}"
even though (\ref{density}) does not describe a wave).
Our convention for Fourier transforms is:
\begin{equation}
\delta_{\vec{k}} = \frac{1}{V} \int \delta_c(\vec{x}, t_0) 
\exp(-i \vec{k} \cdot \vec{x}) d^3 \vec{x}
\label{deltax}
\end{equation}
\begin{equation}
\delta_c (\vec{x}, t_0) = \sum_{\vec{k}} \delta_{\vec{k}} 
\exp(i \vec{k} \cdot \vec{x}) = \frac{V}{(2 \pi)^3} \int \delta_{\vec{k}}
\exp(i \vec{k} \cdot \vec{x}) d^3 \vec{k}
\label{deltak}
\end{equation}
where $V \equiv L_0^3$. 
We take the power spectrum of density fluctuations
(as defined in (\ref{density})
and extrapolated to today in the linear approximation) 
to be of the form
\begin{equation}
P(k)  \equiv  \frac{V}{(2 \pi) ^3} | \delta_{\vec{k}} |^2 
 =  \frac{A w^n}{(1+\eta w + w^2)^2}
\label{P}
\end{equation}
with $w \equiv k/k_{eq}$.
Note that $P(k) \propto k^n$ at $k \ll k_{eq}$
and $P(k) \propto k^{n-4}$ at $k \gg k_{eq}$. 
The scale invariant Harrison-Zel'dovich spectrum 
has $n = 1$.
The power spectrum $P(k)$ is normalized such that
\begin{equation}
\frac{1}{V} \int_{V} \delta_c^2(\vec{x}, t_0) d^3 \vec{x} 
= \int P(k) d^3 \vec{k}
= \sum_{\vec{k}} | \delta_{\vec{k}} | ^2.
\label{normalization}
\end{equation}

\begin{table}
\begin{center}
\begin{tabular}{|rr|rr|}\hline
$\Omega_0$ & $\Delta(\Omega_0)$ & $\Omega_0$ & $\Delta(\Omega_0)$ \\
\hline
$0.2$ & $-1.645$ & $1.0$ & $0.000$ \\
$0.3$ & $-1.066$ & $1.2$ & $0.104$ \\
$0.5$ & $-0.517$ & $1.5$ & $0.216$ \\
$0.7$ & $-0.239$ & $2.0$ & $0.341$ \\
\hline
\end{tabular}
\end{center}
\caption{Function $\Delta(\Omega_0)$
for zero cosmological constant.}
\label{Delta}
\end{table}

\begin{table}
\begin{center}
\begin{tabular}{|rr|rr|}\hline
$\Omega_0$ & $\omega(\Omega_0)$ & $\Omega_0$ & $\omega(\Omega_0)$ \\
\hline
$0.2$ & $0.620$ & $1.0$ & $1.000$ \\
$0.3$ & $0.679$ & $1.2$ & $1.078$ \\
$0.5$ & $0.783$ & $1.5$ & $1.189$ \\
$0.7$ & $0.875$ & $2.0$ & $1.364$ \\
\hline
\end{tabular}
\end{center}
\caption{Function $\omega(\Omega_0)$ for zero
cosmological constant.}
\label{omega}
\end{table}

Equation (\ref{density}) corresponds to the linear 
approximation valid when $\delta_c \ll 1$. 
Exact solutions can be obtained at peaks of the
density fluctuations (assuming peaks of
spherical symmetry and negligible pressure).
When $\delta_c = 1.06$ in the linear approximation
(which has already broken down),
the exact solution is $\delta = 4.55$ and the density 
fluctuation has reached maximum expansion.
When $\delta_c = \delta_m = 1.69$ in the linear approximation,
the exact solution $\delta$ diverges, the density fluctuation
has collapsed and, in our model, a galaxy has formed.

\begin{figure}
\begin{center}
\vspace*{-6.5cm}
\scalebox{0.6}
{\includegraphics[0in,0.5in][8in,9.5in]{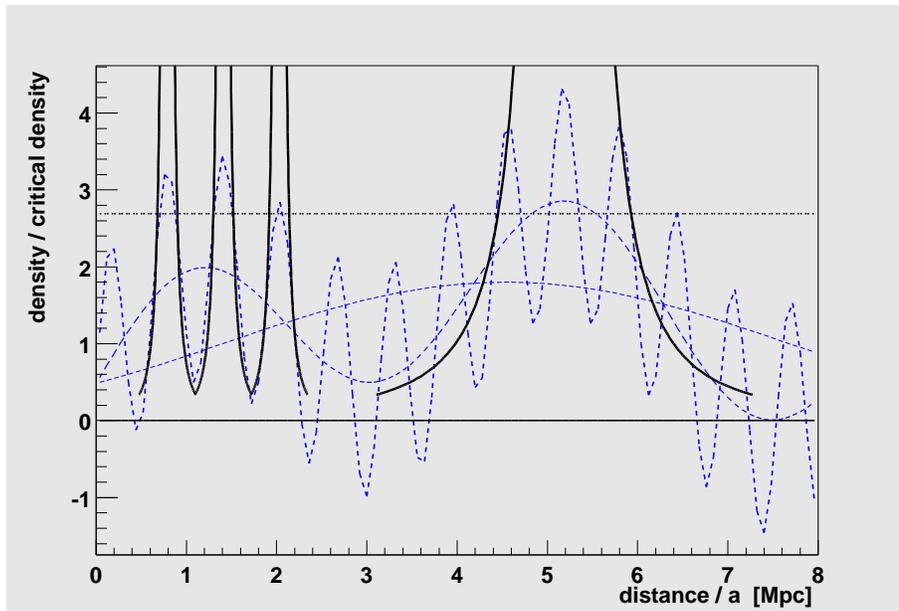}}
\vspace*{0.7cm}
\caption{The hierarchical formation of galaxies.
Dashed lines:
Sum of one, two and three Fourier 
components of the density in the linear approximation.
When the density in the linear approximation 
reaches the dotted line a galaxy forms.
Full lines:
Halos of several galaxies with density run
$\rho \propto r^{-2}$.
As larger wavelengths reach the dotted line,
larger galaxies form 
\textquotedblleft{swallowing}" galaxies of 
previous generations. The peculiar 
displacements of the galaxies have not yet been
applied.}
\label{hierarchy.fig}
\end{center}
\end{figure}

The generation of galaxies on a computer
at a given expansion 
parameter $a_c$ proceeds as follows (see Figure
\ref{hierarchy.fig}). We begin with the largest possible 
galaxy with $I = 2$. We scan the cube searching 
for maximums of $\delta_c(\vec{x}, t)$.
If the maximum at $\vec{x}$ is not \textquotedblleft{occupied}"
by a galaxy (see below), and if 
$\delta_c(\vec{x}, t) \ge \delta_m = 1.69$,
we generate a new galaxy at $\vec{x}$ with the
characteristics listed below. We increase 
$I \rightarrow I + 1$ and repeat the calculation to form 
galaxies of a smaller generation, and repeat this
process up to $I = I_{max}$ corresponding to the
smallest galaxies that we wish to generate.

Note that as time goes on, larger and larger distance
scales $\lambda_I = 2 \pi a_c /k_I$ become non-linear
and new galaxies of larger mass form, 
\textquotedblleft{swallowing}" up galaxies of previous 
generations. Study Figure \ref{hierarchy.fig} again. Galaxy formation is
therefore an ongoing hierarchical process: today's
galaxy clusters are the seeds of tomorrow's
galaxies.

\begin{figure}
\begin{center}
\vspace*{-4.5cm}
\scalebox{0.4}
{\includegraphics[0in,1in][8in,9.5in]{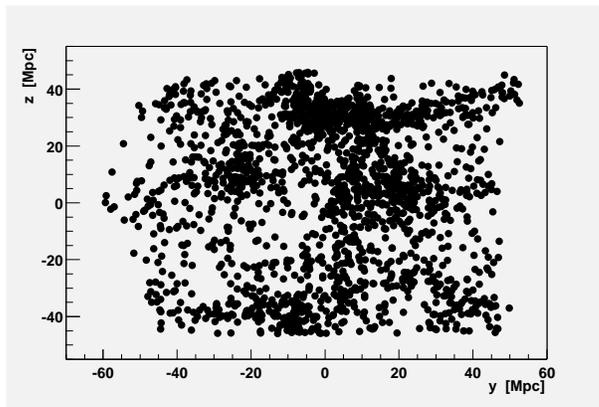}}
\vspace*{0.7cm}
\caption{Scatter plot of $y$ \textit{vs.} $z$.}
\label{yz.fig}
\end{center}
\end{figure}

\subsection{Galaxy characteristics}
We assume that the barionic matter discipates energy and falls
to the bottom of the halo potential wells with
density run $\rho \propto r^{-2}$\cite{Hoeneisen}.
The galaxies generated at \textquotedblleft{generation}"
$I$ have the following properties
\textbf{regarded as approximate descriptions of complex phenomena}. 
Luminous and total
(luminous $+$ dark) radius:
\begin{equation}
\frac{\Omega_0}{\Omega_{lum}}R_{lum} = R = \frac{\lambda_I}{2} 
= \frac{\pi a_c(t)}{k_I},
\label{radius}
\end{equation}
(see Figure \ref{hierarchy.fig}); luminous and total mass:
\begin{equation}
\frac{\Omega_0}{\Omega_{lum}} M_{lum} = M 
\approx \frac{4}{3} \pi R^3 \Omega(t) \rho_c(t), 
\label{mass}
\end{equation}
velocity of circular orbits
(if spiral):
\begin{equation}
v_c = \sqrt{\frac{GM}{R}},
\label{vc}
\end{equation}
or 3-dimensional velocity dispersion 
(if elliptical)\cite{Hoeneisen}:
\begin{equation}
v_{rms} \equiv \sqrt{\langle v^2 \rangle} 
\equiv \sqrt{3} \sigma
= \sqrt{\frac{3GM}{2R}},
\label{vrms}
\end{equation}
peculiar velocity\cite{Weinberg}:
\begin{equation}
\vec{v}_{pec}(\vec{x}, t) \approx \sum_{\vec{k},0 < k \le k_{I-1}}
\frac{2i \vec{k} a_c^2 \vert \delta_{\vec{k}} \vert}{3 k^2 t} 
\cdot exp \left[ i \frac{\vec{k} \cdot \vec{x}}{a_c} + 
i \varphi_{\vec{k}}\right],
\label{vpec}
\end{equation}
(the sum is over $\vec{k}$ up to $k = k_{I-1}$ while
the peculiar motion given by (\ref{xpec}),
$|\vec{x}_{pec}|$, is less than 
$\lambda / 2 = \pi a_c / k$);
and peculiar acceleration:
\begin{equation}
\vec{g}_{pec}(\vec{x}, t) \approx \frac{\vec{v}_{pec}(\vec{x}, t)}{t}.
\label{acceleration}
\end{equation}

What do we mean by \textquotedblleft{occupied}"?
To generate a galaxy of total radius $R$ at position $\vec{x}$
we require that the distance from $\vec{x}$ to already 
generated galaxies of radius $R'$ be greater than
$R' + 0.7R$. The factor $\approx 0.7$ was included to
approximately fill space.

Finally, after generating all galaxies, we correct
their positions $\vec{x}$ by their peculiar motions
\begin{equation}
\vec{x}_{pec} \approx
\int_{0}^{t} \vec{v}_{pec}(\vec{x}, t') \frac{a_c(t)}{a_c(t')} dt'
\approx \frac{3}{2} \vec{v}_{pec}(\vec{x}, t) t .
\label{xpec}
\end{equation}
This step is necessary in order to obtain the
galaxy-galaxy correlation. This completes the 
presentation of the model.

\begin{figure}
\begin{center}
\vspace*{-4.5cm}
\scalebox{0.4}
{\includegraphics[0in,1in][8in,9.5in]{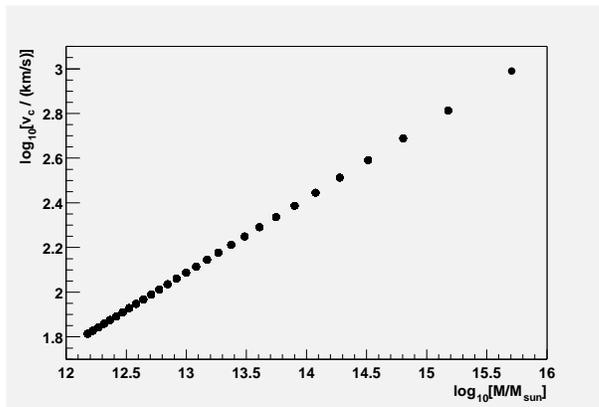}}
\vspace*{0.7cm}
\caption{Scatter plot of $M$ \textit{vs.} $v_c$ to test
the Tully-Ficher relation.}
\label{tully_fisher.fig}
\end{center}
\end{figure}

\subsection{The simulations}
Our simulations are limited by computer resources
(one $500$ MHz, $32$ bit processor). Therefore, we
fix $L_0 = 92$Mpc. Even then the contributions
to $\rho_{lin}$ from Fourier components with
$I = 1$ and edge effects are non-negligible.
More realistic simulations require a larger $L_0$.
We fix $h_0 = 0.6$, $\eta = 2.04$ and
$k_{eq} = 0.155$Mpc$^{-1} \cdot \Omega_0 h_0^2 
\cdot \{3.36/N_{eff}\}^{1/2}$.
This value of $k_{eq}$ corresponds to a distance to the
horizon equal to $\approx 0.5 \lambda_{eq}$ 
at the time $t_{eq}$ when the 
densities of radiation and matter were equal.
$N_{eff}$ is the effective value of $N_b + \frac{7}{8} N_f$,
where $N_b$ ($N_f$) is the number of boson 
(fermion) degrees of freedom. $N_{eff} = 3.36$
for three light neutrino species. 
The values of $k_{eq}$ and $\eta$ were obtained
from a fit to the \textquotedblleft{CHDM}" model
of reference \cite{Silk} which is in agreement with
observations. 

\begin{figure}
\begin{center}
\vspace*{-4.5cm}
\scalebox{0.4}
{\includegraphics[0in,1in][8in,9.5in]{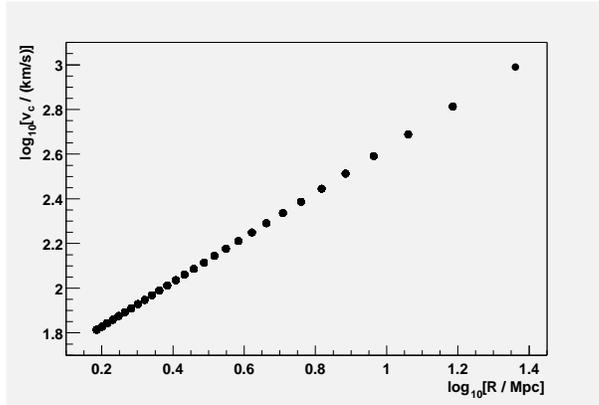}}
\vspace*{0.7cm}
\caption{Scatter plot of $R$ \textit{vs.} $v_c$ to
test the Samurai relation.}
\label{samurai.fig}
\end{center}
\end{figure}

\section{Comparison with observations}
\begin{figure}
\begin{center}
\vspace*{-4.5cm}
\scalebox{0.4}
{\includegraphics[0in,1in][8in,9.5in]{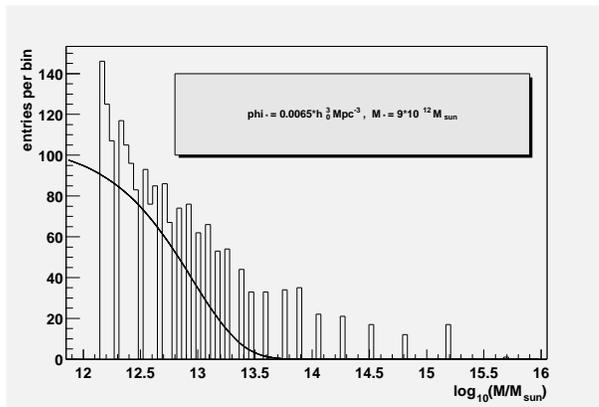}}
\vspace*{0.7cm}
\caption{Histogram of $\log_{10}(M)$ to test the Schechter distribution.}
\label{schechter.fig}
\end{center}
\end{figure}

The results of the model are presented and 
compared with observations in
Figures \ref{yz.fig} to \ref{vy.fig} and 
Table \ref{comparison}.
These figures correspond to the simulation
with $\Omega_0 = 1.0$, $n = 1.00$ and $A = 1706$Mpc$^3$.
Explanations follow.

\subsection{Tully-Fisher, Faber-Jackson and Samurai relations}
The model satisfies $M_{lum} \propto v_c^3 \propto v_{rms}^3$, 
which, if light traces mass, is in
reasonable agreement with the Tully-Fisher relation
for spiral galaxies
($L \propto v_c^\mu$ with $\mu \approx 4$\cite{Peebles}
for the infrared luminosity and $\mu \approx 2.4$
to $2.8$ for the blue luminosity), or the 
Faber-Jackson relation for elliptical galaxies
($L \propto v_{rms}^\mu$ with $\mu = 4 \pm 1$\cite{Peebles}). 
$L$ is the galaxy absolute luminosity.
The model satisfies 
$R_{lum} \propto v_{rms} \propto v_c$, which
is in reasonable agreement with the Samurai relation
for elliptical galaxies ($R_{lum} \propto v_{rms}^\xi$ with
$\xi \approx 1.2$ to $1.3$).
See Figures \ref{tully_fisher.fig} and \ref{samurai.fig}.
Elliptical galaxies satisfy the following
relation with \textquotedblleft{remarkably small}"\cite{Peebles}
scatter: $v_{rms} \propto L^{0.62} R_{lum}^{-0.50}
\propto M_{lum}^{0.50} R_{lum}^{-0.50}$.\cite{Peebles}
This relation is in excellent agreement with our model.

The model satisfies the following useful relations at present:
\begin{equation}
\left( \frac{R_{lum}}{1\textrm{Mpc}} \right)
= 1.4 \Omega_0^{-3/2} \Omega_{lum} h_0^{-1}
\left( \frac{v_c}{100\textrm{km} \cdot \textrm{s}^{-1}} \right),
\label{Samurai}
\end{equation}
\begin{equation}
\left( \frac{M_{lum}}{M_{sun}} \right) = 3.3 \cdot 10^{12}
\Omega_0^{-3/2} \Omega_{lum} h_0^{-1}
\left( \frac{v_c}{100\textrm{km} \cdot \textrm{s}^{-1}} \right)^3,
\label{Tully-Fisher}
\end{equation}
Both equations are in reasonable
agreement with observations if
the density of matter in optically bright baryons is
$\Omega_{lum} \approx 0.005$.

\subsection{The Schechter distribution}
The observed galaxy luminosity distribution is given
by the Schechter rela\-tion.\cite{Peebles} Assuming
that light traces mass, we obtain the
following total (luminous $+$ dark) mass 
distribution of galaxies:
\begin{equation}
\frac{dn}{d\ln(M)} \approx \phi_* 
\exp \left( -\frac{M}{M_*} \right)
\label{Schechter}
\end{equation}
where $dn$ is the number of galaxies per unit volume 
with total mass between $M$ and $M + dM$,
$\phi_* = 0.010 \exp(\pm 0.4) h_0^3$Mpc$^{-3}$
and $M_* = M_{lum*} \Omega_0 \Omega_{lum}^{-1} = 2.8 \cdot 10^{13} 
\exp(\pm 0.4) \Omega_0 h_0^{-1} M_{sun}$.\cite{Peebles}
The corresponding velocity of circular orbits
is $v_{c*}$. See Figures \ref{schechter.fig}
and \ref{vc.fig}.

We choose galaxies of total mass
$M_*$ to be of \textquotedblleft{generation" $I \approx 10$.
Then our simulations require
$L_0 \approx I(6 M_* / \pi \Omega_0 \rho_{c0})^{1/3}$.
We therefore set $L_0 = 92$Mpc as indicated above. 
We also set
$I_{max} = 30$ so that galaxies are generated
down to a mass $\approx M_*/27$. 
Galaxies with less mass contribute
negligibly to the density of the universe and the
computer resources needed to generate them become prohibitive.
The simulations then
probe the power spectrum in the wavevector range
$0.1$Mpc$^{-1} \lesssim k < 2.0$Mpc$^{-1}$.

\begin{figure}
\begin{center}
\vspace*{-4.5cm}
\scalebox{0.4}
{\includegraphics[0in,1in][8in,9.5in]{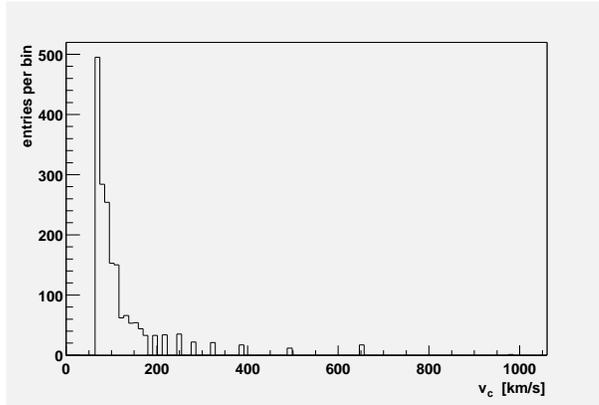}}
\vspace*{0.7cm}
\caption{Histogram of the velocity of circular orbits $v_c$.}
\label{vc.fig}
\end{center}
\end{figure}

\subsection{Galaxy-galaxy correlation}
The observed joint probability of finding two galaxies in two
volume elements separated by $r$ is
\begin{equation}
dP \propto \left[1 + \left( \frac{r_0}{r} \right)^\gamma \right]dV_1 dV_2
\label{correlation}
\end{equation}
with $\gamma = 1.77 \pm 0.04$ and 
$r_0 = (5.4 \pm 1.0)h_0^{-1}$Mpc.\cite{Peebles}
Due to the linear approximation of $\vec{x}_{pec}$
we expect agreement with observations to
break down at small $r$. 
See Figure \ref{correlation.fig}.

\subsection{Fluctuation in galaxy counts}
The observed fluctuation in galaxy counts in randomly
placed spheres of radius $r$ is\cite{Peebles}
\begin{equation}
\frac{ \left< \delta N \right>_{rms}}{\left< N \right>} 
= 1.35 \left( \frac{r_0}{r} \right)^{\gamma / 2}
= 0.84 \pm 0.17
\label{fluctuation}
\end{equation}
for $r=9.2 h_0^{-1}\textrm{Mpc} = 92\textrm{Mpc}/6$ which
is a convenient radius in view of our simulation volume.

\begin{figure}
\begin{center}
\vspace*{-4.5cm}
\scalebox{0.4}
{\includegraphics[0in,1in][8in,9.5in]{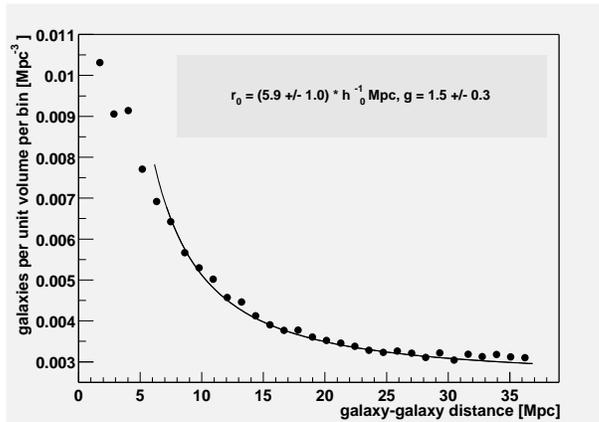}}
\vspace*{0.7cm}
\caption{Histogram of distances between galaxies to test
the galaxy-galaxy correlation.}
\label{correlation.fig}
\end{center}
\end{figure}

\subsection{Fluctuations of the Cosmic Microwave Background}
The Cosmic Microwave Background (CMB) radiation 
propagates freely since matter and radiation decoupled
at $T_{dec} \approx 3000$K, or at
$a_{dec} \equiv (1 + z_{dec})^{-1} \approx 1100^{-1}$.
The relation between the comoving
length at decoupling (or at any time with $a \ll 1$)
with the corresponding angle today is\cite{Peebles}
\begin{equation}
R_0 = \frac{2c}{\Omega_0 H_0} \theta = 
\frac{1.74\textrm{Mpc}}{\Omega_0 h_0} \left( \frac{\theta}{1'} \right)
\label{angle}
\end{equation}
Fluctuations on scales $\theta < 6'$ are erased due to the
thickness of the last scattering surface.
The size of the horizon at decoupling, $3ct_{dec}$, 
corresponds to $\theta_{dec} \approx 1.7^o$.
The size of the horizon at $t_{eq}$, $\approx 2ct_{eq}$,
corresponds to $\theta_{eq} \approx 0.3^o$.

We consider the root-mean-square
fluctuation of the temperature of the CMB on angular
scales $\theta \gg 1.7^o$, \textit{i.e.} fluctuations
that entered the horizon after the decoupling of 
matter and radiation. On these large scales 
\begin{equation}
P(k) = A(k/k_{eq})^n.
\label{P_small_k}
\end{equation}
We use a \textquotedblleft{window function}"\cite{Turner}
\begin{equation}
W(r) = \exp \left( - \frac{r^2}{2 r_W^2} \right)
\label{window}
\end{equation}
which smoothly defines a volume
\begin{equation}
V_W \equiv \frac{4}{3} \pi r_s^3 
= 4 \pi \int_{0}^{\infty} r^2 W(r) dr 
= (2 \pi)^{3/2} r_W^3
\label{volume}
\end{equation}
Then the mean-square fluctuation of mass in
randomly chosen windows of volume $V_W$ is:
\begin{eqnarray}
\left< \frac {\delta M}{M} \right>_{rms}^2 & \equiv &
\frac{1}{V} \int_{V} d^3\vec{x} \left[ \frac{1}{V_W}
\int_{V_W} d^3\vec{r} \delta_c(\vec{x} + \vec{r})
W(\vec{r})
\right] ^ 2 \nonumber \\
& = & \frac{V}{2 \pi^2} \int_{0}^{\infty} 
\left\vert \delta_{\vec{k}} \right\vert^2 
\exp \left( - k^2 r_W^2 \right) k^2 dk
\label{dM}
\end{eqnarray}
at early times, \textit{i.e.} $\Omega \approx 1$.
For the power spectrum (\ref{P_small_k}) in the range
of interest of $n$ we obtain approximately
\begin{equation}
\left< \frac {\delta M}{M} \right>_{rms}^2 \approx
\frac{2 \pi A}{k_{eq}^n r_W^{n+3}}
\label{dMA}
\end{equation}

\begin{figure}
\begin{center}
\vspace*{-4.5cm}
\scalebox{0.4}
{\includegraphics[0in,1in][8in,9.5in]{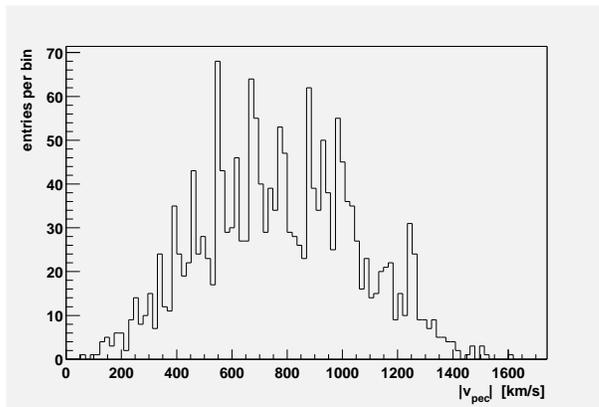}}
\vspace*{0.7cm}
\caption{Histogram of the absolute peculiar velocities of galaxies.}
\label{vpec.fig}
\end{center}
\end{figure}

The fluctuation of the temperature of the CMB
radiation is given by the Sachs-Wolfe 
effect\cite{Peebles,Turner,Longair}. To obtain an
analytic expression we use the prescription:
\begin{equation}
\left<\frac{\delta T}{T} \right>_{rms, \theta} \approx  
\left< \frac{\delta M}{M} \right>_{rms, H, \theta}
\label{Sachs-Wolfe}
\end{equation}
where $\left< \delta M / M \right>_{rms, H, \theta}$
is the root-mean-square fluctuation of mass on the
scale corresponding to angle $\theta$ when that mass 
crossed inside the horizon. 
The prescription (\ref{Sachs-Wolfe}) is in agreement with
numerical calculations\cite{Silk,Peebles}. 
From (\ref{Sachs-Wolfe}), (\ref{dMA}), (\ref{volume}),
(\ref{angle}) for $r_s$, and replacing 
$A^{1/2}$ by the 
amplitude at horizon crossing 
$a_{cH} A^{1/2} \equiv 
a_H f(\Omega_0) A^{1/2} =
\theta ^2 \Omega_0^{-1} f(\Omega_0) A^{1/2}$, 
and substituting numerical values, we finally obtain
\begin{equation}
\left< \frac{\delta T}{T} \right>_{rms, \theta} \approx
\Omega_0^{\frac{1}{2}} \cdot
f(\Omega_0)
\left[ \frac{A}{43\textrm{Mpc}^3} \right]^{\frac{1}{2}}
\left[ \frac{N_{eff}}{3.36} \right]^{\frac{n}{4}}
\left[ \frac{h_0^{n+3-2n}}{598^{n+3} \theta^{n-1}}
\right]^{\frac{1}{2}}
\label{dT}
\end{equation}
with $\theta$ in radians, and 
$f(\Omega_0) = 3(1 - \Omega_0^{-1})/(5 \Delta(\Omega_0))$.
The published fluctuation of the temperature of the CMB is
given in terms of the amplitudes of spherical harmonis
$a_{\ell} \equiv \left< |a_{\ell}^m| \right>$\cite{Peebles}:
\begin{equation}
\left< \frac{\delta T}{T} \right>_{rms, \theta} =
\frac{1}{T_0}
\left[ \sum_{\ell = 2}^{\infty} \frac{2 \ell +1}{4 \pi} 
a_{\ell}^2 \exp \left(-\theta^2 \ell ^2 \right) 
\right]^{\frac{1}{2}}
\label{al}
\end{equation}
for $\theta \ll 2 \pi$.

\begin{figure}
\begin{center}
\vspace*{-4.5cm}
\scalebox{0.4}
{\includegraphics[0in,1in][8in,9.5in]{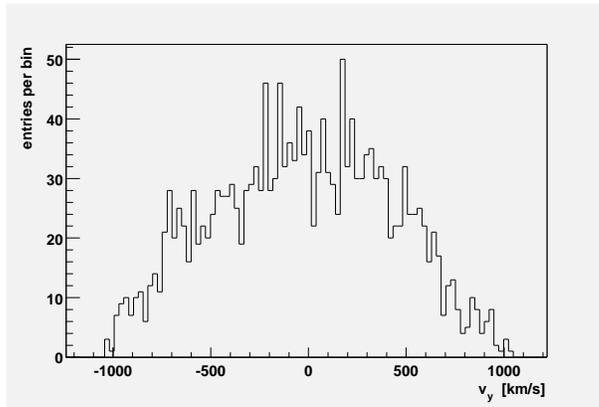}}
\vspace*{0.7cm}
\caption{Histogram of the component $v_y$ of the
absolute peculiar velocity.}
\label{vy.fig}
\end{center}
\end{figure}

\subsection{Peculiar velocities}
The absolute peculiar velocities of galaxies with 
respect to the CMB, shown in
Figures \ref{vpec.fig} and \ref{vy.fig},
have relatively large 
contributions from
wavevectors $k \approx k_{eq}/\eta$ which are well
beyond the range of our simulations. Furthermore, these
absolute peculiar velocities are difficult to measure.
We therefore define a local peculiar velocity
in spheres
of radius $r = 9.2 h_0^{-1}\textrm{Mpc}$:
\begin{equation}
\left< v_{pec} \right>_{rms} \equiv
\left<
\left[ \frac{1}{N} \sum_{i=1}^{N} 
\left| \vec{v}_{pec,i} - \left< \vec{v}_{pec} \right> \right|^2
\right] \right> ^{\frac{1}{2}}
\label{vpecrms}
\end{equation}
where $N$  is the number of galaxies in a sphere
and the average is over all spheres.
This local peculiar velocity can be estimated from the lengths of
the \textquotedblleft{Fingers of God}".\cite{Peebles} See also
\cite{Saslaw}.

\begin{table}
\begin{center}
\begin{tabular}{|c|}
\hline
$\left< \delta T / T \right>_{rms, \theta=0.1745} =
(1.06 \pm 0.04 \pm 0.20) \cdot 10^{-5}$ \\
$v_{c*}  = (220 \pm 50 \pm 60)$km s$^{-1}$ \\
$\left< v_{pec} \right>_{rms} \approx (748 \pm 150 \pm 350)$km
s$^{-1}$ \\
$\delta N / N = 0.84 \pm 0.17 \pm 0.10$ \\
$r_0 = (5.4 \pm 1.0 \pm 2.8)h_0^{-1}$Mpc \\
$\gamma = 1.77 \pm 0.04 \pm 0.54$ \\
$\ln (\phi_* \cdot h_0^{-3} \cdot \textrm{Mpc}^3) = 
-4.6 \pm 0.4 \pm 0.5$ \\
\hline
\end{tabular}
\end{center}
\caption{Benchmark data\cite{Peebles, COBE, Saslaw} 
used to define a $\chi ^2$ to compare
the model with observations. The first error is
observational, and the second one includes
theoretical errors, statistical errors of
the simulations and errors of the fits. 
We add them in quadrature.
All errors
in this article are one standard deviation.}
\label{chisquare}
\end{table}

\subsection{The cosmological constant}
Until now we have set $\Omega_\Lambda = 0$. Let us 
consider a low density spatially flat universe with
$\Omega_0 = 0.3$ and $\Omega_\Lambda = 0.7$. By numerical
integration we obtain
$H_0 t_0 = 0.964$, $\omega = 0.478$, $\Delta = 0.0$,
$\delta_m = 1.85$ and $f = 1.17$. The simulation then
proceeds as before. The results are discussed in the 
next Section.

\begin{table}
\begin{center}
\begin{tabular}{|c|c|c|c|c|c|}
\hline
$\Omega_0$, $\Omega_\Lambda$ $\backslash$ $n$ & $1.50$ & $1.25$ & $1.00$ & $0.75$ & $0.50$ \\
\hline
$0.20$, $0.0$ & $11.8$  & $11.4$  & *  & *  & *  \\
$0.35$, $0.0$ & $72.4$  & $5.7$  & $12.8$  & *  & *  \\
$0.50$, $0.0$ & $196.6$  & $5.8$  & $8.5$  & **  & *  \\
$0.70$, $0.0$ & $24.7$  & $16.2$  & $6.6$  & $11.5$  & *  \\
$1.00$, $0.0$ & $46.3$  & $10.7$  & $3.2$  & $7.3$  & $78.0$  \\
$1.50$, $0.0$ & $162.7$  & $98.4$  & $3.7$  & $5.3$  & $14.3$  \\
$0.30$, $0.7$ & $38.3$  & $9.2$  & $12.2$  & *  & *  \\
\hline
\end{tabular}
\end{center}
\caption{We compare the model with
observations by giving the
$\chi^2$ for $4$ degrees of freedom
for several $\Omega_0$, $\Omega_\Lambda$ and $n$.
The amplitude $A^{1/2}$ was obtained from COBE
data\cite{COBE} and Equation (\ref{dT}) for each
pair $(\Omega_0,n)$. Entries with a star
are excluded because zero galaxies were generated.
Entries with $\chi^2 > 9.5$ are excluded with $95\%$
confidence.}
\label{comparison}
\end{table}

\subsection{Constraints on $\Omega_0$, $n$ and $A$}
To compare quantitatively the predictions of 
the model with observations 
we define a $\chi ^2$ as indicated in Table \ref{chisquare}.
This $\chi ^2$ has seven terms and the model has
three parameters ($\Omega_0$, $n$ and $A$), 
so we are left with four degrees
of freedom. 
We have excluded from $\chi ^2$ the Tully-Fisher and
Samurai parameters $\mu$ and $\xi$,
which are well satisfied, because they are common to 
all simulations.
The theoretical errors of $\delta N/N$, $\gamma$ and $r_0$
quoted in Table \ref{chisquare}
are obtained from an estimate of the error of
$\vec{v}_{pec}$ in the approximation (\ref{vpec}), 
and the propagation of this error 
as determined from numerical simulations.

Several simulations are compared with observations in
Table \ref{comparison}. For each
pair ($\Omega_0$, $n$) we have obtained $A$ from
COBE data (see Table \ref{chisquare}) 
and Equation (\ref{dT}).
Note that we obtain good quantitative
agreement with observations for several pairs
($\Omega_0$, $n$).

The best simulation with $\Omega_\Lambda = 0$ 
has $\chi ^2 = 3.2$, the critical density
$\Omega_0 = 1.0$, the scale invariant
Harrison-Zel'dovich slope $n = 1.00$, and $A = 1706$Mpc$^3$. 
This simulation has $2053$ galaxies,
$M_{lum*} = 9 \cdot 10^{12} M_{sun} \Omega_{lum} \Omega_0^{-1}$,
$v_{c*} = 118$km/s,
$\phi_{*}=0.0065 h_0^3$Mpc$^{-3}$,
$R_{lum*} = 2.8$Mpc$\Omega_{lum} \Omega_0^{-1}$,
$\left< v_{pec} \right>_{rms} = 486$km/s,
$\delta N/N = 0.744$,
$r_0 = (5.9 \pm 1.0) h_0^{-1}$Mpc,
$\gamma = 1.5 \pm 0.3$ (these are errors of the fit),
and a fraction of mass in galaxies $\epsilon = 0.97$.
Several distributions of this simulation are 
presented in Figures \ref{yz.fig} to \ref{vy.fig}.

Some results of the simulations are shown in
Tables \ref{galaxies.tab} to \ref{vcstar.tab}.

\begin{table}
\begin{center}
\begin{tabular}{|c|c|c|c|c|c|}
\hline
$\Omega_0$, $\Omega_\Lambda$ $\backslash$ $n$ & $1.50$ & $1.25$ & $1.00$ & $0.75$ & $0.50$ \\
\hline
$0.20$, $0.0$ & $3017$  & $708$  & $0$  & $0$  & $0$  \\
$0.35$, $0.0$ & $1929$  & $2584$  & $397$  & $0$  & $0$  \\
$0.50$, $0.0$ & $1444$  & $2229$  & $1866$  & $22$  & $0$  \\
$0.70$, $0.0$ & $1146$  & $1765$  & $2285$  & $778$  & $0$  \\
$1.00$, $0.0$ & $890$  & $1398$  & $2053$  & $1937$  & $194$  \\
$1.50$, $0.0$ & $846$  & $1120$  & $1635$  & $2057$  & $1550$  \\
$0.30$, $0.7$ & $1451$  & $2277$  & $1094$  & $0$  & $0$  \\
\hline
\end{tabular}
\end{center}
\caption{Number of galaxies in the simulations 
as a function of $\Omega_0$, $\Omega_\Lambda$ and $n$.}
\label{galaxies.tab}
\end{table}

\section{Conclusions}
We have developed a simple, fast and predictive model of the 
hierarchical formation of galaxies. We obtain
quantitative agreement with observations
(within the limitations of the model, \textit{i.e.}
outside of the core of galaxy clusters). The only free
parameters of the model are $\Omega_0$ and the
power spectrum of density fluctuations,
\textit{i.e.} $n$ and $A$. The COBE observations
determine $A$ as a function of $\Omega_0$ and $n$.
The model provides insight into the hierarchical
formation of galaxies, and
is useful to study the onset of
galaxy formation, the merging of galaxies, 
the redshift-luminosity distributions, and so forth,
and to further constrain $\Omega_0$, $n$, $A$ and $N_{eff}$.

The model is robust for two reasons:
i) Galaxies are not stepped forward in time so errors 
do not accumulate in time; and 
ii) Galaxies are placed where the exact solution for 
the density diverges. 
Weaknesses of the model are:
i) Galaxy formation and merging in the model
is step-wise rather than continuous; and
ii) The peculiar velocities and
displacements are calculated in the linear 
approximation. Therefore, we expect that the 
simple model will not reproduce the center of
galaxy clusters in detail which are in the 
process of merging and have gone non-linear,
nor will it predict correctly the galaxy-galaxy 
correlation at small separation.

Comparison of general galaxy observations with 
simulations determines,
with $95\%$ confidence, that $\Omega_0 > 0.25$ and
$1.1 - 0.3 \Omega_0 < n < 1.4 - 0.2 \Omega_0$
(assuming $\Omega_\Lambda = 0$ and $N_{eff} = 3.36$).
The best simulation has a $\chi^2$ per degree of
freedom equal to $3.2/4$ and corresponds to
$\Omega_0 = 1.0$, $\Omega_\Lambda = 0$ and $n = 1.00$.
A low density flat universe with $\Omega_0 = 0.3$ and
$\Omega_\Lambda = 0.7$ is still allowed. See Table
\ref{comparison} for details.

In summary, we have developed a simple, fast
and powerful tool to study the large
scale structure of the universe.

\begin{table}
\begin{center}
\begin{tabular}{|c|c|c|c|c|c|}
\hline
$\Omega_0$, $\Omega_\Lambda$ $\backslash$ $n$ & $1.50$ & $1.25$ & $1.00$ & $0.75$ & $0.50$ \\
\hline
$0.20$, $0.0$ & $1.06$  & $0.46$  & *  & *  & *  \\
$0.35$, $0.0$ & $1.31$  & $0.76$  & $0.49$  & *  & *  \\
$0.50$, $0.0$ & $1.55$  & $0.93$  & $0.49$  & **  & *  \\
$0.70$, $0.0$ & $1.62$  & $1.15$  & $0.64$  & $0.44$  & *  \\
$1.00$, $0.0$ & $1.37$  & $0.96$  & $0.74$  & $0.50$  & $0.39$  \\
$1.50$, $0.0$ & $1.35$  & $1.40$  & $0.65$  & $0.53$  & $0.38$  \\
$0.30$, $0.7$ & $1.34$  & $0.81$  & $0.45$  & *  & *  \\
\hline
\end{tabular}
\end{center}
\caption{$\delta N / N$ as a function 
of $\Omega_0$, $\Omega_\Lambda$ and $n$. Entries with a
star have zero galaxies.}
\label{dNN.tab}
\end{table}

\begin{table}
\begin{center}
\begin{tabular}{|c|c|c|c|c|c|}
\hline
$\Omega_0$, $\Omega_\Lambda$ $\backslash$ $n$ & $1.50$ & $1.25$ & $1.00$ & $0.75$ & $0.50$ \\
\hline
$0.20$, $0.0$ & $568$  & $234$  & *  & *  & *  \\
$0.35$, $0.0$ & $916$  & $428$  & $204$  & *  & *  \\
$0.50$, $0.0$ & $1086$  & $594$  & $279$  & **  & *  \\
$0.70$, $0.0$ & $1254$  & $761$  & $382$  & $192$  & *  \\
$1.00$, $0.0$ & $1531$  & $967$  & $486$  & $284$  & $130$  \\
$1.50$, $0.0$ & $1668$  & $1085$  & $645$  & $397$  & $220$  \\
$0.30$, $0.7$ & $756$  & $337$  & $156$  & *  & *  \\
\hline
\end{tabular}
\end{center}
\caption{$\left< v_{pec} \right>_{rms}$ in \textrm{km/s}
as a function of $\Omega_0$, $\Omega_\Lambda$ and $n$.
Entries with a star have zero galaxies.}
\label{vpec.tab}
\end{table}

\begin{table}
\begin{center}
\begin{tabular}{|c|c|c|c|c|c|}
\hline
$\Omega_0$, $\Omega_\Lambda$ $\backslash$ $n$ & $1.50$ & $1.25$ & $1.00$ & $0.75$ & $0.50$ \\
\hline
$0.20$, $0.0$ & $43$  & $33$  & *  & *  & *  \\
$0.35$, $0.0$ & $69$  & $60$  & $48$  & *  & *  \\
$0.50$, $0.0$ & $80$  & $80$  & $69$  & **  & *  \\
$0.70$, $0.0$ & $97$  & $97$  & $85$  & $77$  & *  \\
$1.00$, $0.0$ & $118$  & $118$  & $118$  & $109$  & $97$  \\
$1.50$, $0.0$ & $143$  & $129$  & $139$  & $131$  & $143$  \\
$0.30$, $0.7$ & $67$  & $58$  & $53$  & *  & *  \\
\hline
\end{tabular}
\end{center}
\caption{Velocity of circular orbits of $L_*$ galaxies, 
$v_{c*}$, in \textrm{km/s}
as a function of $\Omega_0$, $\Omega_\Lambda$ and $n$.
Entries with a star have zero galaxies.}
\label{vcstar.tab}
\end{table}


\begin{thebibliography}{7}
\bibitem{Weinberg}
S. Weinberg, \textquotedblleft{Gravitation and
Cosmology}", Wiley (1972)
\bibitem{Hoeneisen}
B. Hoeneisen, Jornadas en Ingenier\'{\i}a El\'{e}ctrica
y Electr\'{o}nica, \textbf{12}, 62, Escuela Polit\'{e}cnica
Nacional, Quito, Ecuador (1990);
B. Hoeneisen and J. Mej\'{\i}a, Polit\'{e}cnica 
\textbf{18}, No. 4, 39, Quito, Ecuador (1993);
B. Hoeneisen, CIENCIA, Vol. 1, No. 1, 105, Quito, Ecuador (1998);
B. Hoeneisen, \textquotedblleft{Thermal physics}", 
The Edwin Mellen Press(1993)
\bibitem{Silk}
Eric Gawiser and Joseph Silk, \textit{Science},
\textbf{280}, 1405 (1998). 
Our conventions for $P(k)$ differ by a factor $(2 \pi)^3$.
\bibitem{Peebles}
P.J.E. Peebles, \textquotedblleft{Principles
of Physical Cosmology}", Princeton University Press
(1993)
\bibitem{Turner}
Michael S. Turner, \textquotedblleft{Cosmology
and Particle Physics}" in 
\textquotedblleft{Intersection between Elementary Particle
Physics and Cosmology}", Volume 1, edited by T. Piran and
S. Weinberg, World Scientific (1984)
\bibitem{Longair}
M.S. Longair, in \textquotedblleft{The Deep Universe}"
edited by A.R. Sandage, R.G. Kron and M.S. Longair,
Springer (1995)
\bibitem{COBE}
C.L. Bennett, A. Banday, K.M. Gorski, G. Hinshaw, P.
Jackson, P. Keegstra, A. Kogut, G.F. Smoot, 
D.T. Wilkinson, E.L. Wright, Astrophys.J. 464:L1-L4 (1996) 
\bibitem{Saslaw}
Somak Raychaudhury and William C. Saslaw,
\textquotedblleft{The observed distribution function
of peculiar velocities of galaxies}", 
Spires preprint astro-ph/9602001 (1996)

\end{thebibliography}
\end{document}